# Hybrid Tunable Magnet Actuator: Modeling and Design

W.B. Hoekwater[1,*], J.D. Wiersema[1,*], S.H. HosseinNia[1,**]

[1]Department of precision and microsystems engineering, Delft University of Technology, The Netherlands

[*]Authors equally share contributions

[**]Corresponding author, s.h.hosseinniakani@tudelft.nl

**Abstract**

Reluctance actuators are preferred for high-precision applications. Due to resistive losses in the coils, the accuracy of this type of actuator will reduce in quasi-static operation mode within a vacuum environment. By using soft permanent magnets whose magnetization states are in-situ tuned, no constant power is needed to create a force and thereby the resistive losses will reduce. In this paper, a new tuning method is investigated in order to reduce the resistive losses. By using a history-dependent and non-linear hysteresis model, a more efficient tuning algorithm is designed. Besides this, the position accuracy and control simplicity of a variable reluctance tunable magnet actuator are improved by linearizing the non-linear force-flux relationship. This is achieved by using bias fluxes generated by hard permanent magnets.

## 1. Introduction

In the High-Tech industry, thermal stability within motion and alignment systems is important to realize high accuracy and repeatability. Here vacuum environments and quasi-static operation modes (large bias forces required over "longer" times) are often required, which results in a more critical thermal stability. Two examples of such a system are gravity compensators [1] and deformable space mirrors [2]. Nowadays, reluctance actuators are used to actuate these kinds of systems. These actuators use a coil to generate a flux-regulated force in a magnetic circuit. However, in quasi-static operation, resistive losses in the coil heat up the system reducing the thermal stability [3].

To reduce the resistive losses a constant reluctance actuator was developed that uses the flux of a soft permanent magnet (PM) to create a force. By in-situ tuning the magnetization state (MS), the force can be adjusted [4].

PM-based variable reluctance actuators can improve the application range of such actuators. The schematic shown in Figure 1 illustrates such a concept, a so-called tunable magnet actuator (TMA) that uses a soft PM as a flux source [5]. Current pulses through the coil wound around the PM can tune the PMs' MS to any desired value smaller than the saturated MS to control the actuator force. The realized force-flux relation is non-linear, limiting motion accuracy and control simplicity. The MS is tuned by saturating the magnet and demagnetizing to the desired MS, referred to as the saturation magnetization state tuning (SMST) method. Saturating the PM requires high currents increasing the energy dissipation during the MS tuning. These drawbacks lead to the following developments improving the TMA concept:

a) elimination of the saturation step from the tuning process
b) a linearized force-flux TMA

This paper proposes the concepts for these improvements and their experimental validation.

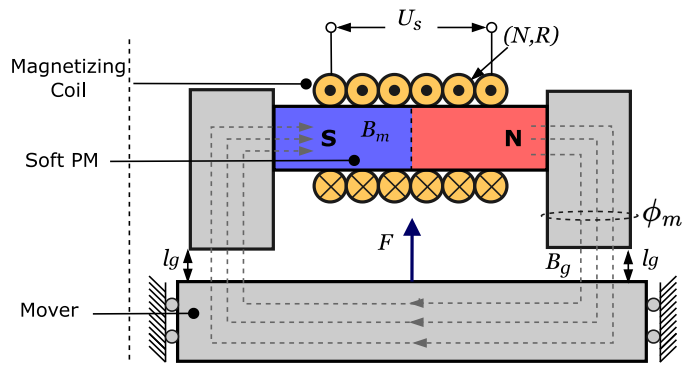

Figure 1: Schematic of a C-shaped tunable magnet actuator

## 2. Continuous magnetization state tuning

To tune a soft PM to a new MS, the corresponding corner point (CP) must be determined. The CP is defined as the point in the B(H) curve where the applied field direction is switched during the tuning process. The SMST calculates the CP on the descending branch of the major loop of the B(H) curve using a linear model which relates the remanent MSs to their corresponding CPs [5].

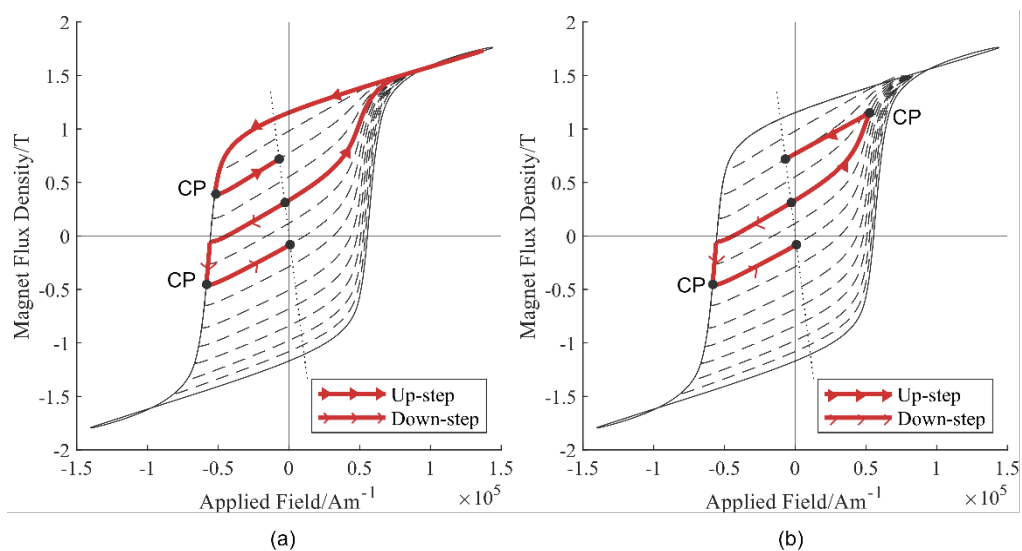

Figure 2: (a) MS tuning using the SMST, where the MS is decreased by the down-step and increased by the up-step. Here, saturating the magnet is needed to increase the MS. b) MS tuning using minor loop reversal curves eliminates the saturation step from the tuning process.

The strategy of only using the descending major loop branch requires returning to the major loop before demagnetizing to the CP, which results in the tuning paths illustrated in Figure 2a. Here the tuning process is described using three steps, namely:

1. Saturation: Saturate the magnet using a large positive applied field to return to the major loop. (only necessary when increasing the MS)
2. Demagnetization: Demagnetize the magnet along the major loop using a negative applied field to reach the calculated CP.
3. Decay: Set the applied field to zero and let the MS decay to its remanent value on the reversal curve.

Saturating the magnet is not needed when decreasing the MS due to the loop closure of the minor loop.

The saturation step can be eliminated from the tuning process by extending the model to include minor loop reversal curves, removing the need to return to the major loop before reaching the CP, illustrated in figure 2b. However, this requires a history-dependent and non-linear CP calculation model [6]. The history of the MS is determined by the loop closure and the wiping out of the property of magnetic hysteresis. An algorithm uses these properties to track the MS history and feed this to the hysteresis model. The algorithm compares the desired remanent MS to the previous states and changes the history accordingly (e.g. in Figure 3, tuning from state 3 to 4 removes state 3 from the history). The non-linear and history-dependent model uses measurement data to calculate the CP based on the history and the desired MS. The algorithm and model facilitate continuous MS tuning without a saturation step as illustrated in Figure 3. Here an arbitrary sequence of MSs and the tuning path used to reach these is shown. This method is referred to as the envelope magnetization state tuning (EMST) method.

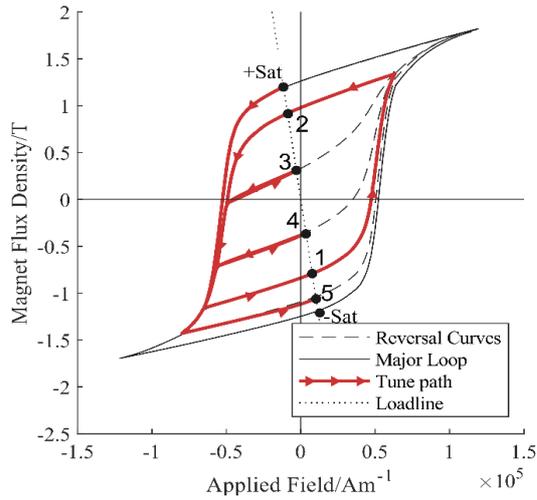

Figure 3: Tuning path following a sequence of MSs using the EMST, eliminating the saturation step by using minor loop reversal curves.

## 3. Linearized force flux TMA

In hybrid reluctance actuators, biased fluxes are used to linearize a reluctance actuator's force-flux relationship [7]. In this actuator, a mover is used that has an air gap on the side of the motion direction. Control flux flows through the mover and air gaps in one direction, while biased flux flows in the opposite direction. Thus, the bias fluxes will linearize the control flux in relation to the force.

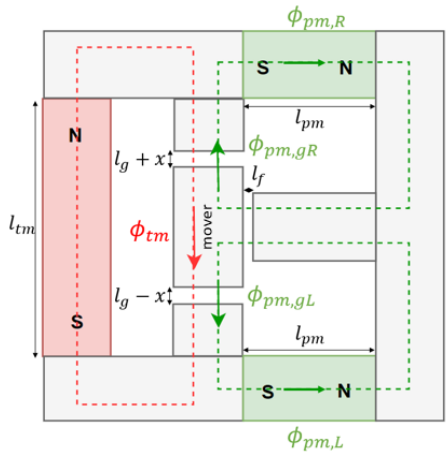

Figure 4: HTMA concept

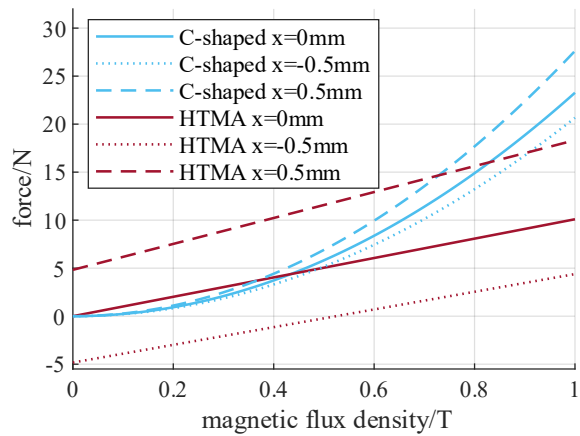

Figure 5: Force-flux relationship for different mover positions

This principle of linearizing can be used for a TMA, which will result in the concept of a hybrid tunable magnet actuator (HTMA) as shown in figure 4 [8]. The control flux is created by a soft PM (red), while the biased flux is created by two hard PMs (green). Using finite element analysis, the behavior of the actuator with a motion range of 500um is characterized in figure 5. It shows the linear force-flux relationship of the HTMA for different positions and compares it with the non-linear relationship of a C-shaped TMA.

By using a piece-wise linear model, the actuator can be described by the linear function:
$$F = K_m \, \phi_{tm} + k_a \, x$$
where $K_m$ is the motor constant, $\phi_{tm}$ is the control flux of the tunable magnet, $k_a$ is the stiffness of the actuator, and $x$ is the position of the mover.

## 4. Experimental validation

The aim of the experimental validation is to analyze the force-flux relationship of an experimental setup and compare it with the FEA analysis and validate the reduced heat generation of the EMST. The schematic actuator of Figure 1 is built, and a force, flux, and position sensor are added to characterize the behavior of the setup, see Figure 6. Figure 7 shows the measured force-flux relation of the experimental setup. Again, the linearity per position is visible as already shown by the FEA. Table 1 compares the SMST to the EMST on the root mean square error (RMSE) of reached MSs and the theoretical average heat dissipation during a tuning step $E_{tune}$. Both methods are tested on twenty random sequences of ten MSs ranging from -1 T to 1 T. The results show a substantial (78%) reduction in heat dissipation with a slight (21%) increase in MS tuning error.

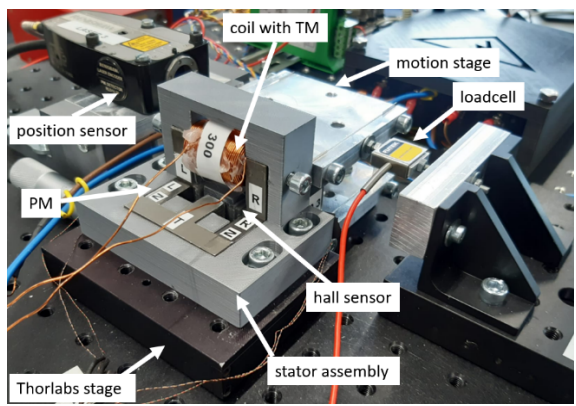
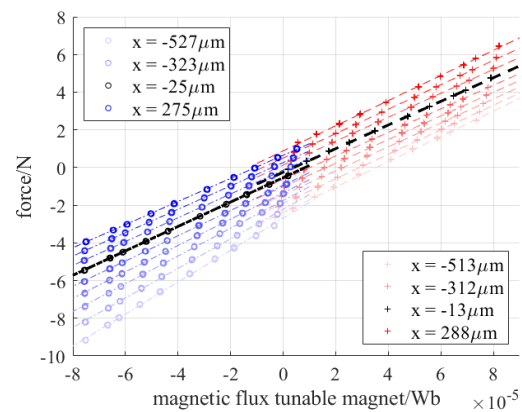

Figure 6: Used experimental setup

Figure 7: Force-flux relationship of the experimental setup

|  | SMST | EMST |
|---|---|---|
| RMSE | 4.8 mT ± 0.8 mT | 5.8 mT ± 0.9 mT |
| $E_{tune}$ | 1.25 J | 0.27 J |

Table 1: Validation of the reduced energy dissipation

## 5. Conclusions

In this paper, a new tuning algorithm based on a data-driven non-linear and history-dependent hysteresis model is designed for variable reluctance tunable magnet actuators. This algorithm removes the saturation step, resulting in an experimentally validated substantial reduction in heat dissipation.

The control accuracy and simplicity of the TMA are improved by changing the actuator design such that a linearized force-flux relationship is obtained. Specifically, bias fluxes are used for this, and the linear relationship is experimentally validated.